\documentclass[aps,prl,twocolumn,10pt,showpacs,superscriptaddress,amsmath,amssymb,nobibnotes]{revtex4-1}

\usepackage{graphicx}
\usepackage{amsfonts}    
\usepackage{graphicx}   
\usepackage{verbatim}   
\usepackage{color}      
\usepackage{subfigure}  
\usepackage{hyperref}   
\usepackage{natbib}
\usepackage{booktabs}
\usepackage{threeparttable}
\usepackage{dcolumn}
\usepackage{multirow}

\begin{document}
\title{Photonic Integrated Quantum Key Distribution Receiver for Multiple Users}
\author{Lingwen Kong}
\altaffiliation{These authors contributed equally}
\affiliation{State Key Laboratory of Optoelectronic Materials and Technologies and School of Physics, Sun Yat-sen University, Guangzhou 510006, China}

\author{Zhihao Li}
\altaffiliation{These authors contributed equally}
\affiliation{State Key Laboratory of Optoelectronic Materials and Technologies and School of Physics, Sun Yat-sen University, Guangzhou 510006, China}

\author{Congxiu Li}
\affiliation{State Key Laboratory of Optoelectronic Materials and Technologies and School of Physics, Sun Yat-sen University, Guangzhou 510006, China}

\author{Lin Cao}
\affiliation{School of Electronics Engineering and Computer Science, Sun Yat-sen University, Beijing 100871, China}

\author{Zeyu Xing}
\affiliation{State Key Laboratory of Optoelectronic Materials and Technologies and School of Physics, Sun Yat-sen University, Guangzhou 510006, China}

\author{Junqin Cao}
\affiliation{State Key Laboratory of Optoelectronic Materials and Technologies and School of Physics, Sun Yat-sen University, Guangzhou 510006, China}

\author{Yaxin Wang}
\affiliation{State Key Laboratory of Optoelectronic Materials and Technologies and School of Physics, Sun Yat-sen University, Guangzhou 510006, China}

\author{Xinlun Cai}
\email{caixlun5@mail.sysu.edu.cn}
\affiliation{State Key Laboratory of Optoelectronic Materials and Technologies and School of Electronics and Information Technology, Sun Yat-sen University, Guangzhou 510006, China}

\author{Xiaoqi Zhou}
\email{zhouxq8@mail.sysu.edu.cn}
\affiliation{State Key Laboratory of Optoelectronic Materials and Technologies and School of Physics, Sun Yat-sen University, Guangzhou 510006, China}
\date{\today}


\begin{abstract}
 
Integrated photonics has the advantages of miniaturization, low cost, and CMOS compatibility, and it provides a stable, highly integrated, and practical platform for quantum key distribution (QKD). While photonic integration of optical components has greatly reduced the overall cost of QKD systems, single-photon detectors (SPDs) have become the most expensive part of a practical QKD system. In order to circumvent this obstacle and make full use of SPDs, we have designed and fabricated a QKD receiver chip for multiple users. Our chip is based on time-division multiplexing technique and makes use of a single set of SPDs to support up to four users’ QKD. Our proof-of-principle chip-based QKD system is capable of producing an average secret key rate of 13.68 kbps for four users with a quantum bit error rate (QBER) as low as 0.51\% over a simulated distance of 20 km in fiber. Our result clearly demonstrates the feasibility of multiplexing SPDs for setting QKD channels with different users using photonic integrated chip and may find applications in the commercialization of quantum communication technology.

\end{abstract}

\maketitle

\section{Introduction} 
Quantum key distribution (QKD) provides absolute security for communication between distant parties, guaranteed by the fundamental laws of quantum mechanics. During the past decades, tremendous progress has been made in experimental demonstration of QKD, which has been realized with optical fibers\cite{MW18,ZQ06} and in free space\cite{B92}, using different degree of freedoms of photons, including polarization\cite{LC10,TL14}, time-bin\cite{CL16,KL15,TY14}, phase\cite{IW02,PM19} and energy-time\cite{AB07,NW13}. 

Recently, integrated photonic technology comes into play in QKD research, and a series of photonic chips designed for QKD applications have been fabricated and used to realize various chip-based QKD experiments, such as coherent one-way QKD \cite{SK17,SE17}, high-dimension QKD with multicore fiber \cite{DB17}, measurement-device-independent QKD \cite{SS20}, pass-block QKD \cite{DZ20} and continuous-variable QKD \cite{ZH19}. In these chip-based experiments, different encodings are employed, including polarization \cite{SK17,MS16,BL19}, path \cite{DB17}, time-bin \cite{SK17,SE17,GZ19} and quadrature \cite{ZH19}. By integrating the optical components of QKD system on a photonic chip, the stability of the QKD system is improved, and its cost may be largely reduced. In this framework single-photon detectors (SPDs), which can be either InGaAs avalanche photodiodes or superconducting nanowire SPDs, have become the most expensive part of an integrated QKD system. In order to fully exploit SPDs and to further reduce the cost of QKD system in view of its commercialization, it is thus necessary to circumvent this issue.

In a QKD network with multiple nodes, a QKD channel between two nodes may be established when one of the two nodes has a transmitter for sending photons and the other node has a receiver able to detect single photons. Typically, there is a one-to-one correspondence between the set of transmitters and that of receivers. On the other hand, if a receiver is able to support multiple transmitters, then SPDs can be multiplexed, thus greatly reducing the cost of a QKD network.

In this letter, we describe a novel QKD receiver photonic chip able to support multiple transmitters. Upon exploiting time-division multiplexing technique, our chip uses a single set of SPDs to establish multiple QKD channels with four different users. We use our chip to implement a proof-of-principle QKD experiment with an average secret key rate of 13.68 kbps for four users and a QBER as low as 0.51\% over a simulated distance of 20 km in fiber.

\begin{figure*}[htbp]

   \includegraphics[scale=0.495]{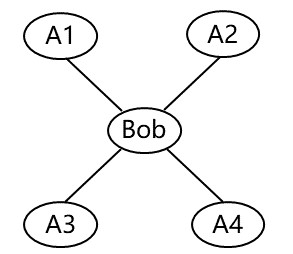}
   \includegraphics[scale=0.495]{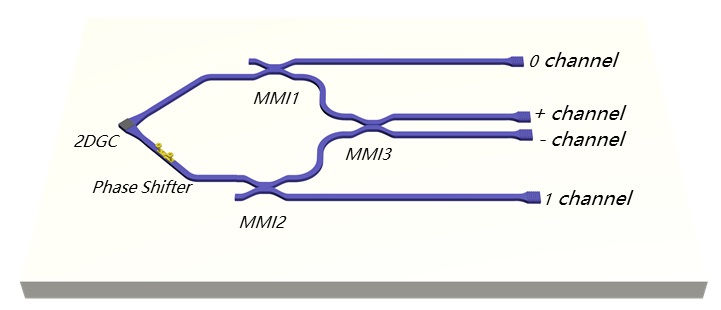}
   \includegraphics[scale=0.495]{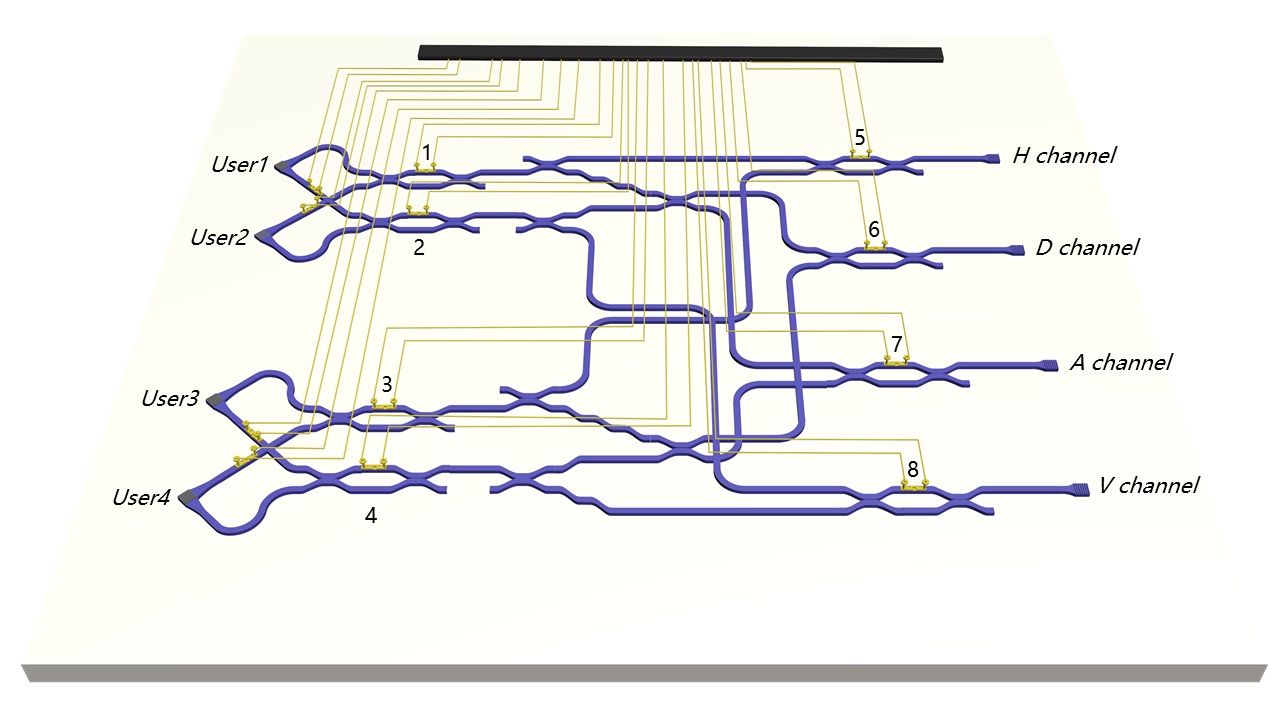}
   \caption{
   The protocol of time-division multiplexing quantum key distribution and the chip fabrication. (a) The four-to-one QKD network. (b) The schematic diagram of one-to-one receiver chip. (c) The schematic diagram of our four-user QKD receiver.
   }
\end{figure*}

\section{The Chip Fabrication} Figure 1a illustrates a QKD network, where Bob wants to establish QKD channel with four different users, A1, A2, A3 and A4. In principle, Bob would need four sets of SPDs to implement QKD with four users. Here we design and fabricate a QKD receiver chip using a single set of SPDs to accomplish the same task. The schematic diagram of our chip is shown in Fig.1c. To explain the functionality of this one-to-four receiver chip, let us briefly review how a one-to-one receiver chip works. Figure 1b shows the schematic diagram of a one-to-one receiver chip. The incident light is coupled into the chip by a two-dimensional grating coupler (2DGC), which is an 18*18 photonics crystal array made of small holes with a diameter of 370 nm. The array converts the encoding photonic quantum state from polarization to path degree of freedom by sending light to the upper or lower arm of the 2DGC when the polarization of the incident photon is horizontal, H, or vertical, V, respectively.  When the polarization of the incident photon is  diagonal, D (anti-diagonal, A), i.e. a superposition of H and V polarizations with a relative phase 0 ($\pi $), the state is transformed into a superposition state of the photon in the upper and lower rail with a relative phase 0 ($\pi $). The photon then goes through three multi-mode interferometers (MMIs). MMI1 and MMI2 are used to passively select the measurement basis: If the photon takes the first or the fourth path, it means that the measurement basis is 0/1; otherwise, the measurement basis is +/-. MMI3 is used to convert the quantum state of a photon in a superposition of path 2 and 3 with a relative phase 0 ($\pi $) to a photon in path 2 (3). Finally, the photon is coupled out into the fibers by the four grating couplers and eventually detected by four SPDs. If the first(second/third/fourth) SPD clicks, it means that the measurement result is 0 (1/+/-). 

Let us now go back to our one-to-four receiver chip shown in Fig. 1c. There are four 2DGCs on the leftmost side of the chip, and each one is connected to a user through an optical fiber. Although four users are connected to this chip, at any one time, only one user can carry out quantum communication through the chip. In order to let the photons of the specific user get through and, at the same time, block the photons from other users, 8 Mach-Zehnder interferometers (MZIs) are used to guide photons to the designated routes. An MZI, composed of one thermo-optic phase shifter and its surrounding two MMIs, serves as an optical switch by applying appropriate voltage on its phase shifter to make the phase shift 0 or $\pi$. Specifically, by setting the phase shifts of MZI1, MZI2, MZI5-MZI8 to be $\pi$, User 1's photons can get through the receiver chip and go to the SPDs. Analogously, to communicate with User 2, the phase shifts of MZI1 and MZI2 should be set to 0 and the phase shifts of MZI5-MZI8 should be set to $\pi$ and to communicate with User 3 (User 4), the phase shifts of MZI3 and MZI4 should be set to $\pi$ (0) and the phase shifts of MZI5-MZI8 should be set to 0.

This multiplexing chip is designed in a way that, by properly setting the phase shifts of the 8 MZIs, the chip serves as a standard QKD receiver for a specific user. In order for the chip to work as designed, the extinction ratio of MZIs needs to be high enough to suppress the crosstalk between different users.

\begin{figure*}[htbp]
  \center
   \includegraphics[scale=0.495]{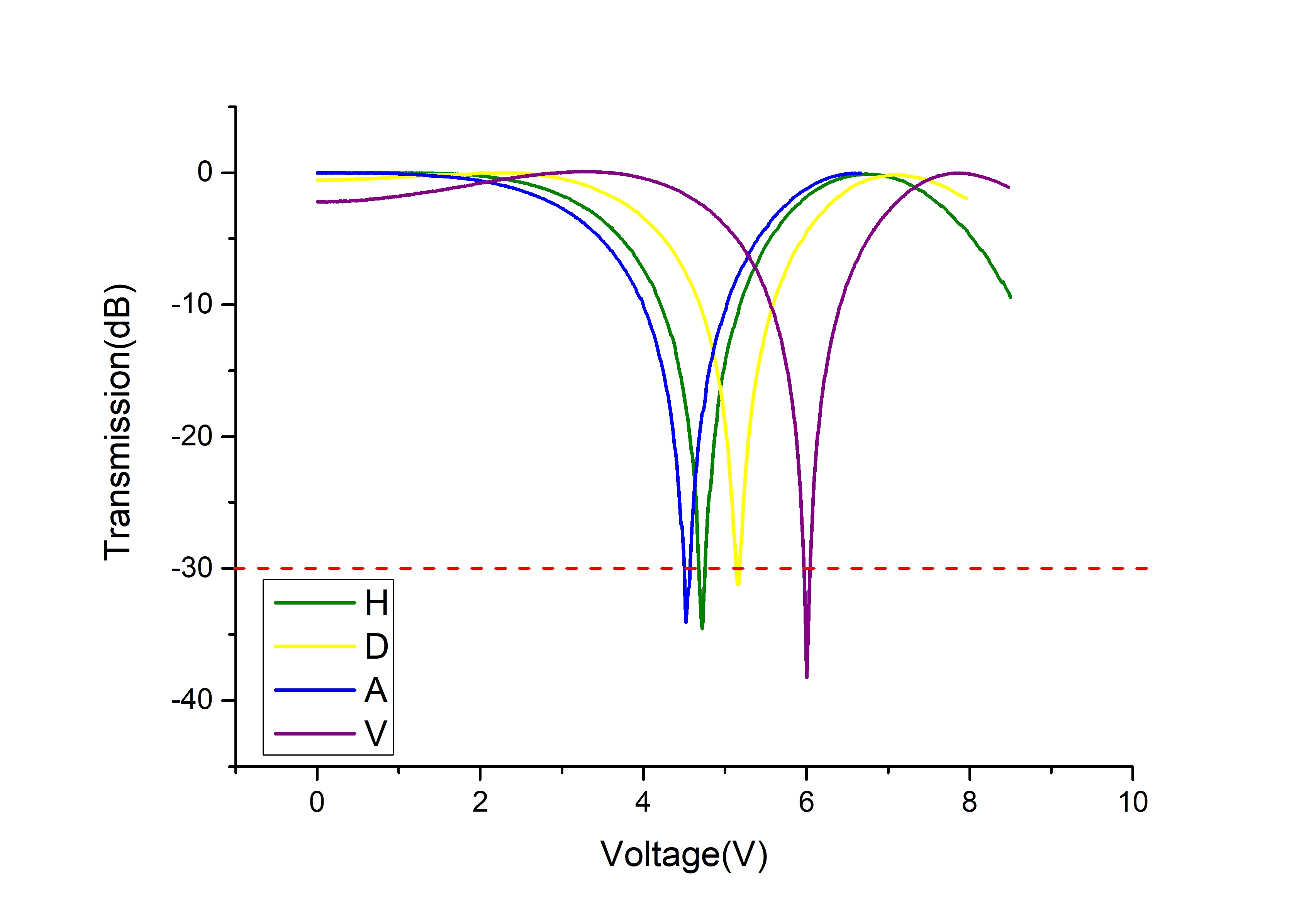}
   \caption{
   The transmission of the MZIs are shown. The extinction ratios are all larger than 30dB.
   }
\end{figure*}

\emph{The extinction ratio of MZIs} - The relative phase between the two arms of the MZI is proportional to the square of the applied voltage. By continually adjusting the voltage, we measured the transmission of the MZIs as a function of the voltages; results are shown in Fig. 2. As it is apparent from the plot, although the extinction ratios of the four MZIs are slightly different, they are all larger than 30 dB. This ensures that the crosstalk between different users is negligible.

\emph{Fabrication process and characterization} - The device was fabricated on a standard 220 nm silicon-on-insulator (SOI) wafer. Firstly, e-beam lithography and inductively coupled plasma etching processes were used to fabricate grating couplers and waveguides. A 1-$\mu$m layer of SiO$_2$ was then coated above the fabricated device as insulation layer. After that, 120-nm Ti was patterned above the specific area as the heaters of MZIs. Finally, 10-nm Ti and 190-nm Au were patterned as the electrodes. After the fabrication of the chip, we then measured its insertion loss at 1550 nm, which is the centering wavelength of the chip. The overall loss of the device is 13 dB, which includes 6-dB loss at the 2DGCs, 5-dB loss at the one-dimensional grating couplers and 2-dB loss from the waveguides of the chip.

\begin{figure*}[htbp]
  \center
   \includegraphics[scale=0.495]{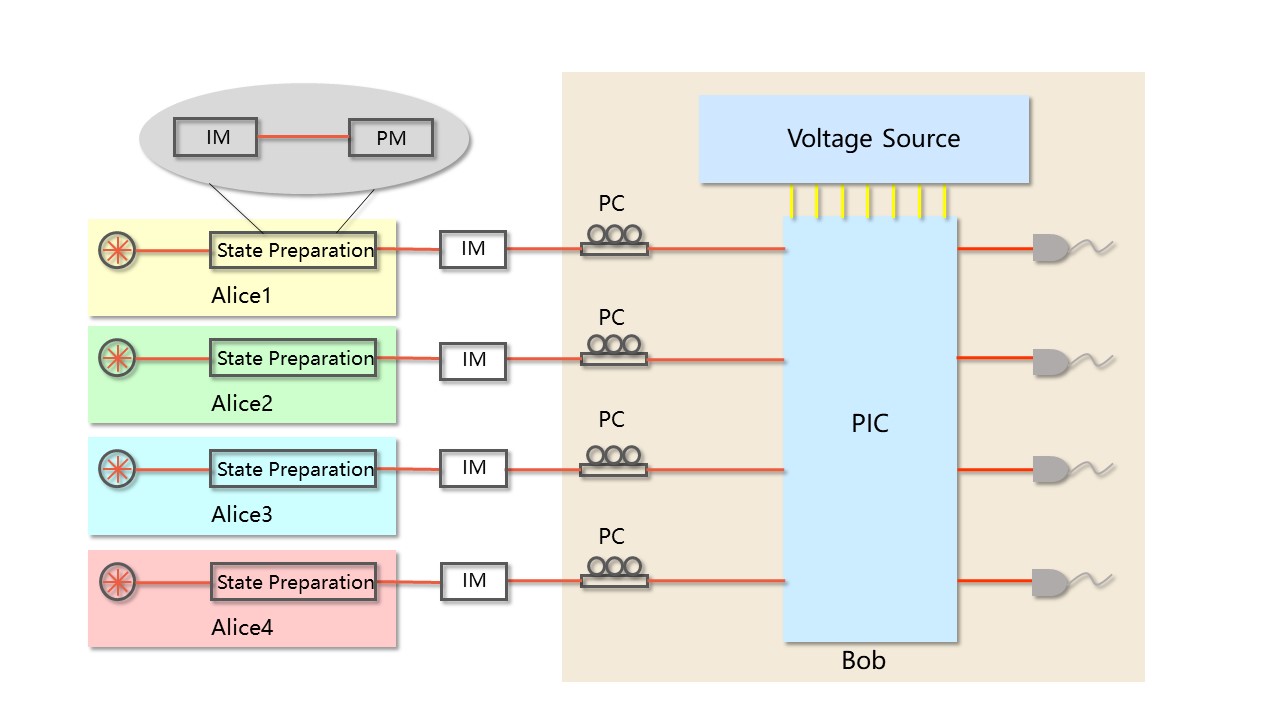}
   \caption{
   The QKD demonstration experimental setup. IM: Intensity Modulator; PM: Polarization Modulator; PC: Polarization Controller; PIC: Photonics Integrated Circuits
   }
\end{figure*}

\section{Chip-based proof-of-principle QKD experiment-} Figure 3 shows the chip-based QKD setup. The output of a pulsed laser with frequency of 10MHz, pulse width of 4 ns, and wavelength 1550 nm passes through an intensity modulator (IM), which is used to attenuate intensity to the single photon level. To implement the decoy state BB84 protocol \cite{W05}, three different intensities of laser pulses are created, with average photon number being 0.6, 0.15 and 0 for the signal and two decoy states chosen with probabilities of 0.5, 0.25, and 0.25 respectively. After this, a polarization modulator (PM), composed of two quarter-wave plates and a half-wave plate, prepares the input light to the desired BB84 polarization state.

The pulses then need to go through an optical channel, which typically is an optical fiber. Here we use a tunable attenuator to simulate a standard optical fiber (0.2 dB/km) with various lengths to reduce the complexity of experiment. We note that, the dispersion caused by a 100-km-long fiber is about 220 ps, which has no effect on a QKD system with a repetition rate of 10 MHz. As a result, it is valid to use an attenuator to simulate the fiber and such approach has been widely used in other works\cite{SK17,SE17,ZH19,GZ19}.

After going through the tunable attenuator, the input pulses are then coupled into the chip through a V-groover fiber array. By setting the phase shifts of the 8 MZIs in the chip, we can choose which one of the four users is able to communicate with. After passing through the chip, the optical pulses are coupled out of the chip through a fiber array to reach the superconducting SPDs. Here the detection efficiency of superconducting SPDs at 1550 nm is $\sim$80$\%$ and the dark count rate is $\sim$120 per second. 

Figure 4 shows the estimated key rates per signal and the QBERs for the four users. Here the error correction efficiency used in the experiment is 1.16 \cite{MQ05}. To reduce the impact of the polarization instability in optical fibers, we shortened the time of data acquisition and used polarization controller (PC) for polarization compensation before running the experiment. Taking into account that the repetition rates of the laser is 10 MHz, we estimate an average secret key rate of 13.68 kbps for four users and a QBER as low as 0.51\% over a simulated distance of 20 km in fiber.

\begin{figure*}[htbp]
   \includegraphics[scale=0.3]{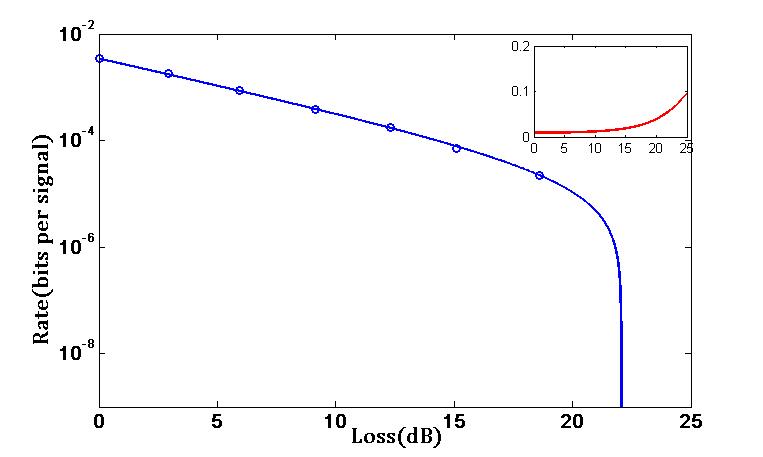}
   \includegraphics[scale=0.3]{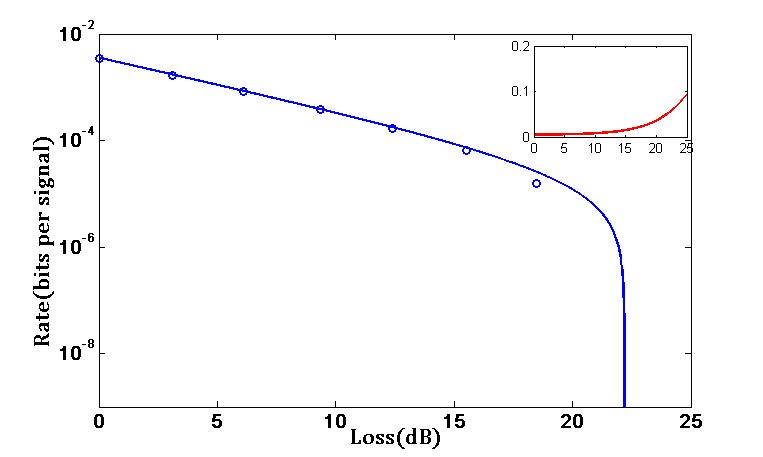}
   \includegraphics[scale=0.3]{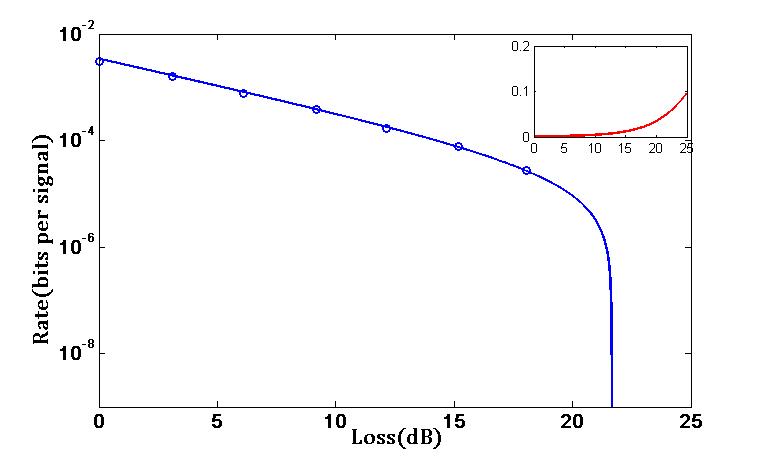}
   \includegraphics[scale=0.3]{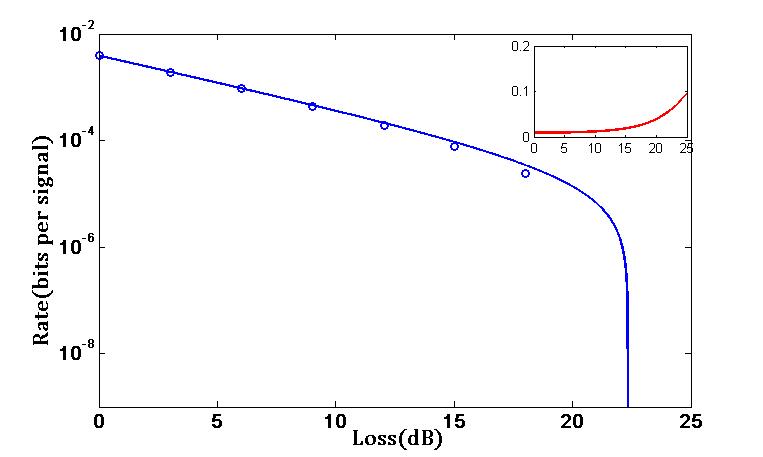}
   \caption{
   The estimated key rates and QBERs for the four users. (a)-(d) corresponds to the estimated key rates and QBERs of Users 1-4. The blue circles represent the measured points in the experiment, and the blue curve is the theoretical estimation. The estimated QBER curve is the red curve in the inset in the upper right corner of each panel. The estimated key rates (per pulse) of the four users are 0.001205, 0.001397, 0.001339 and 0.001529, respectively, whereas the corresponding QBERs are 0.5205\%, 0.5353\%, 0.6185\% and 0.3559\% over 20 km fiber.
   }
\end{figure*}

Due to the non-ideal extinction ratio of the MZIs, when the chip is set to communicate with a specific user, a tiny proportion of the optical signals sending from other users can reach the SPDs. Here we implement a series of experiments to evaluate the influence of such crosstalk on the key rates and the QBERs. The chip is set to communicate with A4 and the communication distance is 20 km. At first, only A4 sends signals to Bob. In this scenario, the key rate (per pulse) is 0.001489 and the QBER is $0.36\%$. We then let both A3 and A4 send signals to Bob. In this case, the key rate becomes 0.001443 and the QBER becomes $0.50\%$. When three users A2, A3 and A4 all send signals to Bob, the key rate becomes 0.001426 and the QBER becomes $0.53\%$. When all four users send signals to Bob simultaneously, the key rate becomes 0.001417 and the QBER becomes $0.56\%$. From the above results, one can see that the decrease of the key rate caused by such crosstalk is less than 5$\%$ and the QBER is well below 1$\%$.

\section{Discussion} Although the present chip is based on polarization encoding, this time-division multiplexing technique is by no means limited to this encoding, but can also be applied to other encodings such as time-bin \cite{SK17,SE17,GZ19}, path \cite{DB17} and quadrature \cite{ZH19}.

We note that the function of the current chip can also be realized by using a standard QKD receiver chip \cite{SK17,SE17,MS16,BL19} combined with an off-chip optical switch. However, by integrating the two functions of optical switch and polarization decoding together into a single device, our chip can reduce both the insertion loss and the complexity of the QKD system.

\section{Conclusion} In summary, we have designed and fabricated a multi-user quantum key distribution receiver chip based on time-division multiplexing. Our proof-of-principle chip-based QKD system is capable of producing an average secret key rate of 13.68 kbps and a QBER as low as 0.51\% over a simulated distance of 20 km in fiber. Our results prove the feasibility of multiplexing SPDs for setting QKD channels with different users using photonic integrated chip, and may find applications in the commercialization of quantum communication technology.

\begin{acknowledgments}
This work was supported by the National Key Research and Development Program (2017YFA0305200 and 2016YFA0301700), the Key Research and Development Program of Guangdong Province of China (2018B030329001 and 2018B030325001), the National Natural Science Foundation of China (Grant No.61974168) and the Natural Science Foundation of Guangdong Province of China (2016A030312012). X.Z. acknowledges support from the National Young 1000 Talents Plan.

The authors faithfully thank Lidan Zhou and Lin Liu for their assistance for the fabrication of the chip.

\end{acknowledgments}

\end{document}